# Simplified model for periodic nanoantennae: linear model and inverse design

**Joshua Borneman\*, Kuo-Ping Chen, Alex Kildishev, and Vladimir Shalaev**

*Electrical and Computer Engineering Department, Birck Nanotechnology Center,
Purdue University, 1206W State Street, West Lafayette, IN 47907
\*Corresponding Author: jdbornem@purdue.edu*

**Abstract:** We determine and use a minimal set of numerical simulations to create a simplified model for the spectral response of nanoantennae with respect to their geometric and modeling parameters. The simplified model is then used to rapidly obtain best-fit modeling parameters to match experimental results, accurately predict the spectral response for various geometries, and inversely design antennae to have a desired performance. This method is structure and model independent, and is applied here to both nanoantenna pair arrays and strips modeled using a 3D finite-element method and 2D spatial harmonic analysis, respectively. Typical numerical simulations may need hours per model, whereas this method, after the initial time to obtain a baseline set of simulations, requires only seconds to analyze and generate spectra for new geometries.

## 1. Introduction

The design and fabrication of optical nanoantenna metamaterials is important for a broad variety of applications, including near-field scanning optical microscopy (NSOM), enhanced Raman scattering, biosensors, sub-wavelength resolution, and nano-scale optical lithography.[1-8] Nanoantennae rely on field enhancement due to plasmon coupling between paired metal nanostructures, such as bow-ties or paired ellipses [3, 6, 9-13], strips [8, 14, 15], rods [1, 5, 16-18], disc [19, 20], or core-shell structures [21-23]. In order to properly design these structures for a specific performance, and for post-fabrication retrieval or optical properties, computationally intensive simulations are normally used, including finite element method [8, 24], finite difference time domain [9, 10, 25], discrete dipole approximations [20, 22, 26], boundary element method [16, 22], and spatial harmonic analysis [27].

The performances of fabricated nanoantennae typically differ from ideal models due to effects not accounted for in the model. The actual size and shape of the antennas differ from the initial design due to fabrication limitations and systemic error. Additionally, the properties of the metal may differ from those used in simulations. For example, the permittivity of gold is typically taken from bulk film measurements, but does not account for gold lattice defects, chemical size effects, and most importantly, surface roughness, which can have a significant effect on nanoparticle resonances [28]. Therefore, in order to match an experimental result with a simulation, the parameters of the simulation must be changed to 'effective' values which account for these discrepancies.[24] Many time consuming simulations may be required to find parameters, which match the experimental result.

Additionally, pre-fabrication simulations are required to determine which dimensions will lead to a desired performance. Inversely designing a nanoantenna using nature-inspired optimization methods [29] can do this, although feedback from fabrication is necessary in order to account for non-ideal fabrication constraints and variations in material properties as explained above, while also being time consuming, especially for a 3D simulation.

In order to reduce the time and computer resources required, we create a simplified equivalent model of a nanoantennae system, which results in the ability to analyze many geometries with minimized computational effort. We apply an equivalent model to both nanoantenna pair arrays and strips using an experimental design methodology[30-34]. This results in a model based on simple linear equations, which accurately describe the antenna system over a limited parameter range, and although illustrated here with paired nanoparticles and strips, may be generally applied to any simulation.

## 2. Nanoantennae fabrication and modeling

The paired nanoantennae diagramed in Fig. 1 (first published as Sample A in [24], but herein identified as Sample 1) were used in addition to the nanostrips shown in Fig. 2. The nanostrips were fabricated on a glass substrate coated with 15 nm of indium-tin-oxide (ITO). E-beam lithography was used with a PMMA resist to pattern the nanoantennae. The substrate was coated with 5 nm of titanium as an adhesion layer followed by the periodic gold nanoantennae array. In Sample 1, the nanoantenna pair array is the array of paired elliptic-shape gold particles, and sample 2, nanoantenna strips, consists of a periodic array of paired gold strips. Both of these structures are resonant in optical frequencies. Transmission and reflection spectra over the visible range are taken using linearly polarized light; incident polarization across the gap induces the primary resonance ($\hat{x}$ direction; P polarization), and polarization parallel to the gap induces a secondary resonance in the case of pair arrays, or no resonance in strips ($\hat{y}$ direction; N polarization).

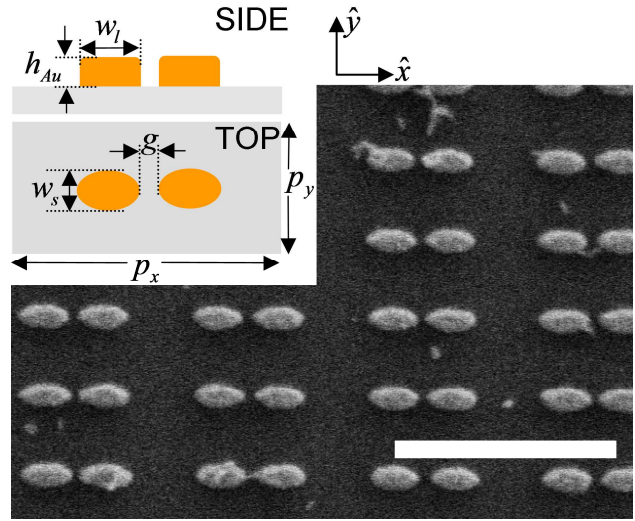

Fig. 1. Nanoantennae sample 1 SEM (sample A in [24]) [scale bar = 500 nm]. Inset represents 3D FEM geometry.

Sample 1, nanoantenna pairs were modeled using finite-element method multiphysics (FEM) through commercially available Comsol Multiphysics software. The general model consist of 6 geometric dimensions: gap, $g$, Au height, $h_{Au}$, long-axis, $w_l$, short-axis, $w_s$, x-period

($p_x$), y-period ($p_y$) and a variable Drude loss factor [24, 28] for gold ($\alpha$) to account for surface roughness. Sample 2 (see Fig. 2) is modeled using a 2D spatial harmonic analysis solver [27] (SHA), which is possible due to the semi-infinite nature of the strips. The nanoantenna strip model uses 5 geometric variables: gap, $g$, width, $w$, period, $p$, Au antennae height $h_{Au}$, Ti layer height, $h_{Ti}$, and also a variable Drude loss factor for gold, $\alpha$.

Typically, the post-fabrication geometry is found from SEM measurements with limited resolution. More precise effective dimensions are determined by matching modeled transmission and reflection spectra with the experimental results. This 'fabrication-calibrated' model, using the proper effective dimensions, can then be used to retrieve optical properties such as index of refraction, and the magnitude and wavelength dependence of the local field enhancement.

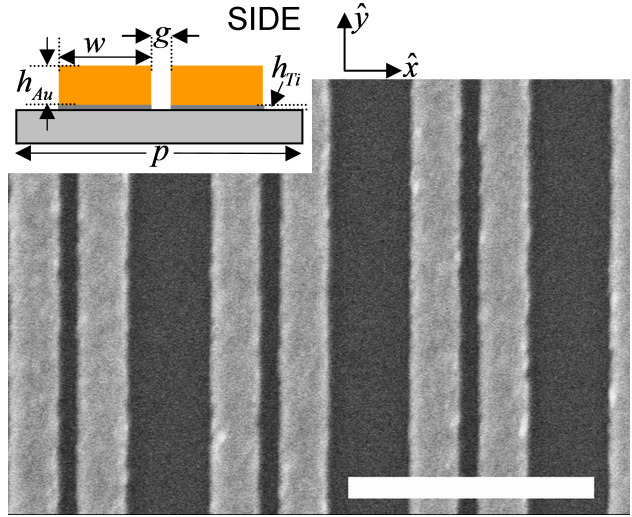

Fig. 2. Nanoantenna Strips sample 2 SEM [scale bar = 500 nm]. Inset represents 2D SHA model geometry.

## 3. Simplified system

We create a set of simulations, which is used to determine a simplified system describing the results. In our case, a linear equation with variable interactions is used, which once defined, may then quickly predict the results of future simulations. The simulation has $n$ input parameters $P = [p_1, ..., p_n]$ (these could be model variables such as width, height, etc.), output $f$, and coefficients $\beta$ relating the two. A general polynomial system is shown (up to quadratic terms) in Eq. (1); more complex higher-order terms may be added if necessary.

$$f(P) = \underbrace{[\beta_0 + \beta_1 p_1 + ... + \beta_{n+1} p_n]}_{linear} + \\ \underbrace{[\beta_{n+2} p_1 p_2 + ... + \beta_{2n} p_1 p_n + \beta_{2n+1} p_2 p_3 + ...... + \beta_j p_{n-1} p_n]}_{interaction} + \quad (1) \\ \underbrace{[\beta_j p_1^2 + ... + \beta_j p_n^2]}_{quadratic} + ...$$

The methodology we follow is a well developed approach called experimental design [34]. This method is widely used to analyze systems with multivariate input using a limited number of experiments, and is typically applied when experiments are costly, such as in biological research, and although the method has existed for some time, its application to metamaterial electromagnetic modeling has not been explored. Herein we use a two-level fractional-factorial [30-33] experimental design.

In order to use this system to predict the output $f$, we must first determine $\beta$. The solution to this system may be more readily understood using matrix notation. For the linear plus interaction system considered here, there is one constant term, $n$ linear terms, and $\frac{1}{2}(n-1)n$ interaction terms for a total of $t = \frac{1}{2}(n^2 + n + 2)$ terms. Eq. (1) is rewritten as $\mathbf{X}\beta = \mathbf{f}$, where $\mathbf{X}$ is an $m \times t$ matrix (a set of $t$ input parameters for each of $m$ simulations). $\beta$ is a $t$-dimensional vector of coefficients (a coefficient for each term), and the output $\mathbf{f}$ is an $m$-dimensional vector (a scalar result for each of the $m$ simulations). Note that although mathematically the system input is $\mathbf{X}$, it is determined from the fundamental model parameters $P$, e.g. the $m^{th}$ row of $\mathbf{X}$ is given by:

$$X_m = \begin{bmatrix} 1 & p_{m1} & \cdots & p_{mn} & p_{m1}p_{m2} & \cdots & p_{m1}p_{mn} & p_{m2}p_{m3} & \cdots & p_{m2}p_{mn} & \cdots & p_{m(n-1)}p_{mn} \end{bmatrix}$$

We use a two-level design for $P$, so that over the $m$ simulations each input parameter $p_i$ is either a low or high value, -1 or 1 (e.g. width is 90 nm or 110 nm). In order to get all $t$ terms of vector $\beta$, we must run enough simulations so that each of the $t$ columns of $\mathbf{X}$ are linearly independent, i.e., the system is not under-determined. For instance, if we have 5 input parameters, then $t = 16$ and we must design 16 sets simulation parameters $P_m$, $m = 1\ldots16$, so that we have 16 linearly independent simulations with $X_m$, $m = 1\ldots16$, and thus, $\mathbf{X}$, is a $16 \times 16$ matrix. This is easily done with common statistical programs, or for example with Matlab's 'fracfact' function [35]. With $P_m$ properly defined (which in turn gives $X_m$), we then run the simulations to retrieve the results $f_m$, so we may finally solve for all coefficients $\beta$ using $\mathbf{X}\beta = \mathbf{f}$. With vector $\beta$ obtained, the simplified system is determined. From now on, we may use Eq. (2), for any set of input parameters $P$ (which therefore defines $X$, a $1 \times t$ matrix), to predict the output $f$ (a scalar value).

$$X\beta = f \qquad (2)$$

**4. Simplified nanoantennae model**

Applying experimental design methodology to nanoantennae modeling reduces the number of full simulations that must be run to a reduced set, which may then be used to describe the system. Any number of input variables $n$ may be used; potential input variables for the current samples include geometric parameters such as gap, width, height, and periodicity, or other parameters such as loss factor. Our initial goal is to find the model parameters which result in spectra that best match the experimental data. In this work, we use 3D FEM and 2D SHA simulations, however it is important to note that since the analysis only relies on input parameters $P$ versus an output value $f$, the simulation itself is a 'black box' and may consist of any modeling technique desired. While $f$ may be defined as one of many potential output values, we define it using one of two separate but complimentary methods.

*4.1 Fit parameters for nanoantennae*

In the first method, each set of simulated spectra (transmission and reflection at both polarizations) are defined according to five output values. These values are specifically chosen to describe nanoantennae spectra, and may not apply to other types of models. These values are: $f_1 = \lambda_{EP} - \lambda_{SP}$ (error in resonance wavelength between the simulation and the experimental value for P polarization, $f_2 = A_{EP} - A_{SP}$ (error in resonance magnitude between the simulation and the experimental value for P polarization), $f_3 = \lambda_{EN} - \lambda_{SN}$ (error in resonance wavelength for P polarization), $f_4 = A_{EN} - A_{SN}$, (error in resonance magnitude for N polarization), and $f_1 = W_{EP} - W_{SP}$ (error in full-width-half-max for P polarization). Since the resonance position error is on the order of tens of nanometers, but magnitude error is between 0 and 1, the five output values are normalized to a range from -1 to 1. We therefore have five $f$ output values for each simulation. A separate simplified model (separate set of coefficients $\beta$) is determined independently for each of the five fit criteria, meaning that we can predict each fit parameter for a given input $P$. In order to achieve the overall best fit between the experiment and a simulation, the input values $P$ are found to minimize a weighted root-mean-square of the five fit criteria, $F$, Eq. (3). A weight is ascribed to each of the fit parameters to allow for a difference in the importance of each fitting criterion. Minimizing Eq. (3), where each value $f_i$ is a simple linear equation, results in a rapid determination of the values for the model which best fit the given experiment.

$$F = \sqrt{\sum_{i=1}^{5} \overline{w}_i f_i^2}, \quad \overline{w}_i = w_i \bigg/ \sum_{i=1}^{5} w_i \qquad (3)$$

Using this simplified system of five fit parameters also, and perhaps more importantly, allows us to reverse the process and inversely design the nanoantennae. Instead of providing experimental data to determine each simulation's error, $f_1$ to $f_5$, we may instead provide desired values for an optimal nanoantenna, (e.g. a desired resonance wavelength). In this case minimizing Eq. (3) determines the values for the model, $P$, which result in the desired performance. Using this method, many antennae may be designed rapidly and the results explored, simply through minimizing a simple equation. This results in effective and accurate generation of design parameters for a desired antenna performance.

*4.2 Fit per wavelength*

A second method is to define an individual output $f$ as the percent transmission or reflection for each spectrum at each wavelength. For example, suppose each simulation outputs four spectra (transmission and reflection for P and N), and wavelengths of 400 nm to 900 nm in steps of 20 nm (25 steps), then we would have 100 different $f$ values, and therefore 100 $\beta$ vectors. This complete set of coefficients allow us to recreate or predict the full spectra, comparable to an FEM or SHA model, for any set of input parameters $P$. However, since these spectra are the result of solving simple equations, the spectra are obtained near-instantaneously. This method, similar to using the 'fit parameters' method, may also be used to fit experimental data. Model input values $P$ may be determined which minimizes the root-mean square error between the entire experimental spectra and the predicted spectra. This method has the advantage over the 'fit parameters' method that it is not model specific, and will fit spectra for any model, however it cannot be used for inverse design.

*4.3 Inverse design*

A combination of both methods leads to a complete inverse design system. The performance of the antenna (primary and secondary resonance wavelengths, resonance magnitudes, etc.) may be manually defined, and the 'fit parameters' method may be used to determine the optimal model values, $P$, to achieve this result. Similar to more complex inverse design optimization methods [29], our simple model may be used to find an optimal design, and additionally if the simplified model were built from an array of experimental samples, instead of simulations, the fabrication and material variations that limit stochastic optimization methods would be built inherently into our simplified system, whether or not the deviations were measured, or even understood. After determining an optimized geometry, the 'per wavelength' method may then be employed to generate an entire set of predicted spectra for the inversely designed sample.

## 5. Results

### 5.1 Nanoantennae fitting

The nanoantenna pair array (sample 1)[24] as described in Fig. 1, has dimensions $[g, w_l, w_s, h_{Au}, p_x, p_y]$ of $[20, 111, 57, 40, 400, 200]$ nm, taken from the evaporator's Quartz Crystal Microbalance (QCM) and final SEM image with a typical error of $\pm 7$ nm. This sample has the greatest fabrication error in gap and long and short axis widths, as well as in loss factor ($\alpha$). Therefore, the values of each variable in the two-level experimental design are set to be $[g, w_l, w_s, \alpha] = [(12, 28)\,\text{nm}, (104, 118)\,\text{nm}, (52, 62)\,\text{nm}, (2, 6)]$ With four variables, $n = 4$ necessitating 11 simulations to achieve a resolution adequate to obtain Eq. (1) with two-way interaction terms, other dimensions $[h_{Au}, p_x, p_y]$ were held as constant due to their low relative error.

Table 1. Nanoantennae (sample 1) Coefficients

| X term | Corresponding Model Parameter | $\beta$ for $f_1$ | $\beta$ for $f_2$ | $\beta$ for $f_3$ | $\beta$ for $f_4$ | $\beta$ for $f_5$ |
|---|---|---|---|---|---|---|
| 1 | - | 0.040 | -0.189 | -0.531 | -0.084 | -0.265 |
| $p_1$ | $w_l$ | -0.451 | 0.171 | 0.037 | 0.079 | -0.183 |
| $p_2$ | $w_S$ | 0.170 | 0.048 | -0.201 | 0.457 | -0.008 |
| $p_3$ | $g$ | 0.377 | -0.073 | 0.039 | -0.016 | 0.070 |
| $p_4$ | $\alpha$ | -0.006 | -0.530 | 0.119 | -0.400 | -0.445 |
| $p_1 p_2$ | $w_l \cdot w_S$ | 0.038 | 0.003 | 0.063 | -0.006 | -0.001 |
| $p_1 p_3$ | $w_l \cdot g$ | 0.017 | 0.008 | -0.014 | 0.021 | 0.006 |
| $p_1 p_4$ | $w_l \cdot \alpha$ | -0.006 | -0.007 | -0.016 | -0.023 | -0.019 |
| $p_2 p_3$ | $w_S \cdot g$ | 0.011 | -0.004 | 0.025 | 0.001 | 0.003 |
| $p_2 p_4$ | $w_S \cdot \alpha$ | -0.004 | 0.006 | -0.031 | -0.108 | 0.010 |
| $p_3 p_4$ | $g \cdot \alpha$ | -0.004 | 0.007 | 0.006 | 0.005 | 0.011 |

The resulting linear coefficients $\beta$, for each of the five fit parameters of section 4.1 are shown in Table 1. This table, combined with Eq. (1) or (2) allows one to retrieve the fit parameters for any values of $P = [g, w_l, w_s, \alpha]$ within the two-level range, without actually running a full simulation. The optimized parameters $P_{fit}$ are found by minimizing the total fit $F$, Eq. (3), and leads to best-fit model values $P_{fit}$ of $[g, w_l, w_s, \alpha] = [27\,\text{nm}, 117\,\text{nm}, 57\,\text{nm}, 3.4]$.

The ability of the 'fit per wavelength' method to predict entire spectra was used to generate the spectra for the given best-fit values, and is shown in Fig. 3 compared to the experimental data. In order to validate the ability to predict an entire set of spectra using method 2, the predicted best-fit curves are also compared to actual the FEM models, which were simulated using the above best-fit values. The maximum error between the actual FEM model and the simple linear model prediction spectra averages less than 1%, with a peak of 4% near the resonance. This indicates that a linear model is accurate over the range of parameters used, and agrees well with the predicted spectra.

In addition to achieving a best-fit model to experimental data, the effects of each variable on the response of the antenna may be analyzed. For instance, if we consider the slope of each term of $f_1$ ($\beta$ coefficients in Table 1), the function describing the error in the wavelength of the primary resonance, with respect to each input variable, we may understand which dimensions are essential to the robustness of the design, and to the response of the nanoantennae. We immediately see that, in order of importance, the long-axis and the gap have the greatest effect on the primary (P) resonance location, whereas $f_2$, the resonance magnitude is dominated by the loss factor, $\alpha$, as might be expected.

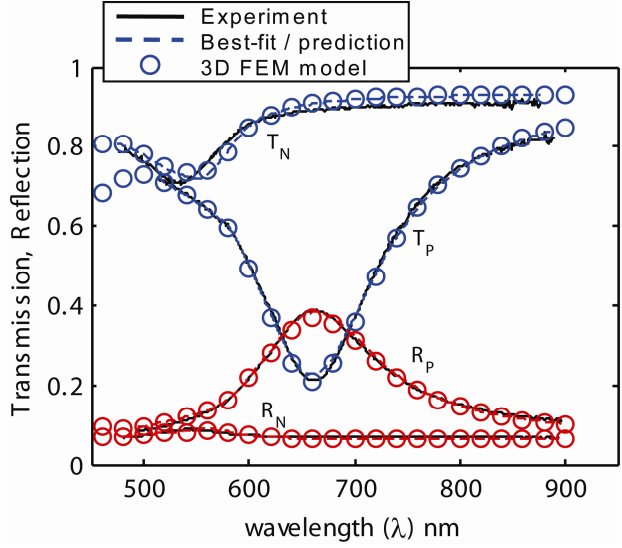

Fig. 3. Nanoantenna array (sample 1), transmission and reflection for primary polarization (P) and secondary polarization (N). Experimental results vs. method 1 (fit parameters) best fit. Predicted spectra were generated using method 2 (fit per wavelength). The results of the actual 3D FEM model are shown using the same best-fit parameters

*5.2 Nanoantennae inverse design*

Instead of using the existing fabricated sample, we also inversely design an optimum nanoantennae structure by manually defining the values $\lambda_{EP}, A_{EP}, \lambda_{EN}, A_{EN}, W_{EP}$ from section 4.1. For instance, instead of getting $\lambda_{EP} = 660$ nm from experimental data, we manually define a desired resonance, such as $\lambda_{EP} = 690$ nm, and then calculate $f_1 = \lambda_{EP} - \lambda_{SP}$ for each model. We can also optimize the strength of the resonance by defining $A_{EP} = 0$, meaning that we would like the primary resonance to be as strong as possible, having ideally zero transmission. Using a set of manually defined values then describes a set of desired spectra, and therefore a desired nanoantenna. Subsequent minimization of equation (3) leads to ideal values for the geometry $[g, w_l, w_S]$ of $[12, 118, 55]$ nm. The predicted spectra show a strong resonance with only 10% transmission at 690 nm as shown in Fig. 4, and the result is closely matched by an FEM model with the same parameters.

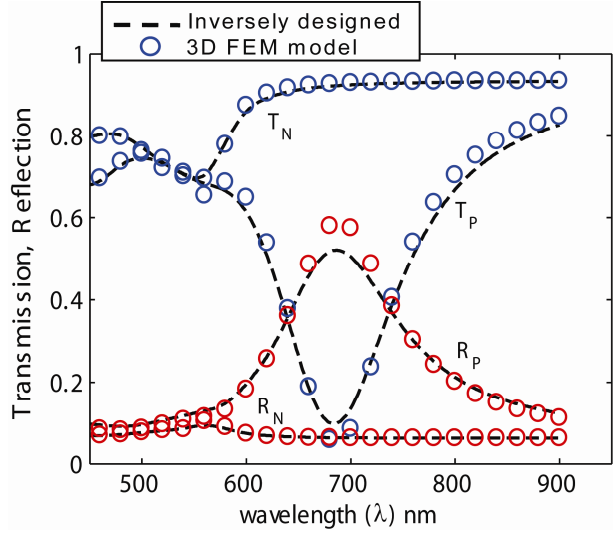

Fig. 4. Predicted transmission and reflection spectra for nanoantenna array geometry inversely designed to have a primary resonance at 690 nm . Predicted spectra versus results of the actual 3D FEM model for the designed values.

*5.3 Nanoantenna strips fitting*

A similar analysis was performed on sample 2, nanoantenna strips, as shown in Fig. 2.

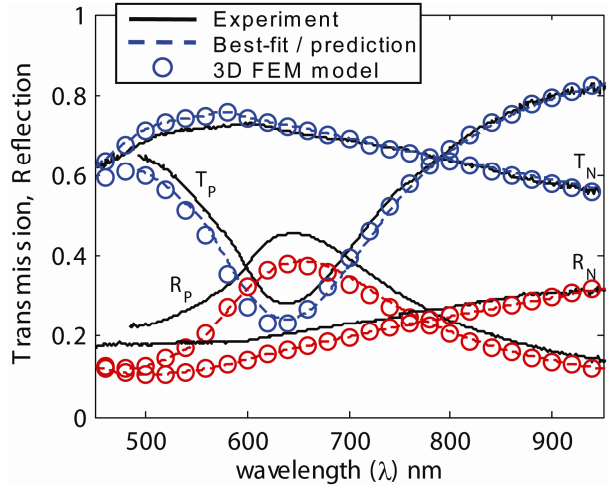

Fig. 5. Nanoantenna Strips (sample 2), transmission and reflection for primary polarization (P) and secondary polarization (N). Experimental results vs. method 1 (fit parameters) best fit. Predicted spectra were generated using method 2 (linear fit per wavelength). The results of the actual 2D SHA model are shown using the same best-fit parameters, and agree well with the predicted curve.

From the QCM and SEM image, the dimensions $[g, w, h_{Au}, h_{Ti}, p]$ are $[38, 98, 20, 5, 400]$ nm respectively, with a typical error of $\pm 5$ nm. This sample has the greatest fabrication error in gap ($g$), width ($w$), Au height ($h_{Au}$), Ti height ($h_{Ti}$); and loss factor[24] ($\alpha$). The values for each variable in the two-level experimental design $[g, w, h_{Au}, h_{Ti}, \alpha]$ are set to be $[(33, 41)\,\text{nm}, (93, 103)\,\text{nm}, (14, 22)\,\text{nm}, (1, 5)\,\text{nm}, (1, 5)]$. 5 variables necessitate 16 simulations to determine the coefficients to Eq. (1) or (2). The periodicity, $p$, was held as a constant due to its low relative. The resulting coefficients are shown in Table 2, and the best-fit results are $[g, w, h_{Au}, h_{Ti}, \alpha] = [40\,\text{nm}, 93\,\text{nm}, 16.4\,\text{nm}, 1\,\text{nm}, 3.2]$. The comparison between the experiment, best-fit predicted spectra and actual 2D SHA solutions for the best-fit values are shown in Fig. 5.

Table 2: Nanoantenna Strips (sample 2) Coefficients

| X term | Corresponding Model Parameter | $\beta$ for $f_1$ | $\beta$ for $f_2$ | $\beta$ for $f_3$ | $\beta$ for $f_4$ | $\beta$ for $f_5$ |
|---|---|---|---|---|---|---|
| 1 | - | 0.127 | -0.115 | -1.000 | 0.409 | -0.337 |
| $p_1$ | $w$ | -0.188 | 0.169 | 0.000 | 0.124 | -0.121 |
| $p_2$ | $g$ | 0.050 | -0.034 | 0.000 | 0.019 | -0.004 |
| $p_3$ | $h_{Au}$ | 0.448 | 0.113 | 0.000 | 0.353 | 0.130 |
| $p_4$ | $\alpha$ | 0.052 | -0.345 | 0.000 | 0.048 | -0.160 |
| $p_5$ | $h_{Ti}$ | 0.163 | -0.307 | 0.000 | 0.094 | -0.240 |
| $p_1 p_2$ | $w \cdot g$ | 0.008 | -0.015 | 0.000 | 0.002 | -0.013 |
| $p_1 p_3$ | $w \cdot h_{Au}$ | 0.017 | 0.000 | 0.000 | 0.014 | -0.001 |
| $p_1 p_4$ | $w \cdot \alpha$ | -0.013 | -0.006 | 0.000 | -0.008 | -0.005 |
| $p_1 p_5$ | $w \cdot h_{Ti}$ | -0.008 | -0.001 | 0.000 | -0.003 | -0.012 |
| $p_2 p_3$ | $g \cdot h_{Au}$ | -0.007 | -0.008 | 0.000 | 0.005 | 0.001 |
| $p_2 p_4$ | $g \cdot \alpha$ | -0.012 | 0.001 | 0.000 | -0.003 | 0.003 |
| $p_2 p_5$ | $g \cdot h_{Ti}$ | -0.016 | 0.001 | 0.000 | -0.003 | -0.004 |
| $p_3 p_4$ | $h_{Au} \cdot \alpha$ | 0.007 | 0.013 | 0.000 | -0.026 | 0.012 |
| $p_3 p_5$ | $h_{Au} \cdot h_{Ti}$ | -0.035 | 0.048 | 0.000 | -0.022 | 0.052 |
| $p_4 p_5$ | $\alpha \cdot h_{Ti}$ | 0.053 | 0.113 | 0.000 | -0.001 | 0.047 |

$f_3$ is zero due to no P resonance.

The error between the predicted spectra and the actual SHA model again averages less than 1%, with a peak error of 6%. An analysis of the function coefficients shows that for nanoantenna strips the position of the primary resonance depends on not only the width of the strips and somewhat on the gap, but is significantly affected by the thickness of the gold and titanium layers. It is also apparent that the titanium layer, in addition to affecting the primary resonance wavelength, also decreases the resonance magnitude with strength similar to the modeling loss factor. Models not including the titanium adhesion layer may result in an unnecessarily large loss factor for gold.

**6. Conclusion**

By using a simplified model with linear and interactions terms, we are able to accurately describe the effects of geometric and modeling parameters on nanoantennae transmission and reflection spectra. This simplified model can be used to quickly fit experimental data with effective model dimensions using a limited baseline set of numerical models. Once this baseline set is obtained, in addition to fitting, we may inversely design nanoantennae to meet specified criteria. In a few moments, the geometry necessary for a specific resonance shift may be determined, without the need for numerous intensive simulations.

This simplified model predicts simulation spectra with an error of less than 4% for both ellipse and strip nanoantennae modeled using 3D finite-element method and 2D spatial harmonic analysis, although it is general to any geometry or simulation method. Further work may be done to determine the range of parameters over which only linear and interaction terms are sufficient. A broader range may lead to inaccurate results due to a nonlinear effect of geometric dimensions on spectra. Alternative higher-order models may be needed for a broader parameter range. However, the simplified model used here is sufficient for experiment matching and inverse design.

**Acknowledgements**

The authors are grateful to Dr. Drachev for useful discussions and Dr. Boltasseva for fabricating Sample 1; partial support from ARO-MURI awards 50342-PH-MUR and W911NF0610283 is highly appreciated.